\documentclass[floatfix,reprint,amsmath,amssymb,aps,10pt]{revtex4-2}
\usepackage{ragged2e}
\usepackage{siunitx}
\usepackage{setspace}

\usepackage{amsmath}
\usepackage[dvipsnames,svgnames,x11names,table]{xcolor}
\usepackage{transparent}
\usepackage{booktabs,tabularx}
\usepackage{graphicx}% Include figure files
\usepackage{dcolumn}% Align table columns on decimal point
\usepackage{bm}% bold math
\usepackage[colorlinks, citecolor = black, urlcolor = RoyalBlue, linkcolor = RoyalBlue]{hyperref}

\begin{document}
\newcommand{\SnV}{\ensuremath{\text{SnV}^-}}
\everymath{\rm} 

\preprint{APS/123-QED}

\title{Tunable cavity coupling of a single \SnV center in nanodiamond across bad-emitter and bad-cavity regimes}

\author{S. Sachero\texorpdfstring{$^1$}{1}, R.Berghaus\texorpdfstring{$^1$}{1}, E. Nieto Hernandez\texorpdfstring{$^2$}{2}, F. Feuchtmayr\texorpdfstring{$^1$}{1},  N. Lettner\texorpdfstring{$^1$}{1}, P. Maier\texorpdfstring{$^1$}{1}, S. Ditalia Tchernij\texorpdfstring{$^2$}{2}, A. Kubanek\texorpdfstring{$^1,^3$}{1,3}}

\email{alexander.kubanek@uni-ulm.de}

\affiliation{$^1$Institute for Quantum Optics, Ulm University, 89081 Ulm, Germany}

\affiliation{$^2$Physics Department, University of Torino, via P. Giuria 1, 10125 Torino, Italy}

\affiliation{$^3$ Center for Integrated Quantum Science and Technology (IQst), Ulm University, Albert-Einstein-Allee 11, 89081, Ulm, Germany}

\begin{abstract}

Efficient coupling between quantum emitters and optical cavities is essential for scalable quantum photonic technologies. Group IV vacancy centers in diamond, particularly the negatively charged tin-vacancy center, have emerged as promising candidates due to their spectral stability, high Debye–Waller factor and large orbital splitting in ground-states. Here, we demonstrate controlled coupling of a single negatively charged tin vacancy center hosted in a nanodiamond to a fully tunable Fabry–Perot microcavity. At cryogenic temperatures, we access the weak coupling regime and observe a transition from the bad-emitter to the bad-cavity regime as the optical transition of the color center narrows. At 4 K, a Purcell factor exceeding 1.7 is measured, confirming cavity-enhanced emission. The obtained results demonstrate the potential of \SnV centers in nanodiamonds as a coherent single-photon source for quantum networks.

\end{abstract}

\maketitle 

\section{\label{sec:level1}Introduction}
Coupling a quantum emitter to a single mode of light is fundamental for various quantum technologies, such as quantum repeater and quantum networks \cite{Kimble2008, Hucul2015, quantum_network}. A key requirement for the successful implementation of these applications is the generation of coherent single photons \cite{doi:10.1126/science.aam9288, rogers2014all}, which must be emitted in a well-defined optical mode. The efficiency of coherent photon emission primarily depends on the intrinsic properties of the emitter. In this context, group IV vacancy (G4V) centers in diamond have emerged as promising candidates \cite{IWASAKI2020237, PhysRevB.109.085414, chen2020building}. The aforementioned centers exhibit a high Debye–Waller factor, with over 60\% of their emission residing within the zero-phonon line (ZPL) \cite{harris_hyperfine_2023, dHaenens-Johansson2011, Iwasaki2015, PhysRevB.107.214110}. Furthermore, their $D_{3d}$ symmetry renders them robust against electric field fluctuations \cite{stark_effect, Becher_2014, Rugar2020}, ensuring stable and coherent emission. Among these centers, the negatively charged tin-vacancy (SnV$^-$) center stands out due to its high quantum efficiency \cite{Iwasaki2017, PhysRevB.99.205417, tchernij_single-photon-emitting_2017} and a large ground-state (GS) orbital splitting ($\sim$850 GHz) \cite{Gali_2018}, which significantly suppresses phonon-mediated decoherence at moderate cryogenic temperatures.

To further enhance photon emission into the ZPL, color centers can be integrated into optical microcavities. Depending on the atomic polarization decay rate ($\gamma$), cavity-field decay rate ($\kappa$) and atom-cavity coupling rate (g), different regimes of cavity quantum electrodynamics (cQED) can be accessed. From the strong coupling regime ($g \gg \kappa, \gamma$) \cite{tolazzi_continuous_2021}, where coherent energy exchange dominates \cite{PhysRevLett.98.117402, hennessy_quantum_2007}, to the weak coupling regime, where loss processes ($\gamma$ or/and $\kappa$) dominate the atom-cavity coupling rate \cite{fox_optical_2012, Saavedra:21}.

%the cavity modifies the spontaneous emission rate, leading to enhanced emission via the Purcell effect \cite{fox_optical_2012, Saavedra:21}. 
In the weak coupling regime, two distinct sub-regimes can be identified depending on the relative magnitudes of $\gamma$ and $\kappa$: the bad-emitter regime ($\kappa \ll \gamma$) \cite{PhysRevLett.110.243602}, where the emitter’s rapid dephasing rate dominates the emitter-cavity interaction, and the bad-cavity regime ($\kappa \gg \gamma$) \cite{PhysRevLett.112.043601}, where the cavity decay rate is the fastest decay channel.
In the bad-emitter regime, the cavity funnels the emission into a narrower cavity mode \cite{riedel2017deterministic, FP_NV_RT}, thereby yielding a frequency-tunable bright and narrow single-photon source \cite{PhysRevLett.110.243602, PhysRevApplied.13.064016}. Further narrowing of the emitter linewidth, achievable, for example, through cryogenic cooling in solid-state defect, can bring the system into a transitioning regime where $\gamma < \kappa$ \cite{PhysRevLett.112.043601}, which in this paper we refer to as bad-cavity regime. In this regime, the cavity modifies the spontaneous emission rate, leading to enhanced emission via the Purcell effect \cite{ruf2021resonant, yurgens_cavity-assisted_2024, PhysRevApplied.7.024031}.

The coupling efficiency is limited by competing loss channels. In particular, unwanted scattering or absorption processes, commonly denoted as $\kappa'$, contribute to additional losses that degrade the overall system efficiency. These losses compete with the desired cavity-field decay rate $\kappa$, effectively reducing the fraction of photons emitted into the target optical mode. As a result, only a fraction of emitted photons is channeled into the well-defined cavity mode. Therefore, minimizing $\kappa'$ is critical for optimizing the performance of cavity-enhanced single-photon sources.

Open-access Fabry–Perot (FP) microcavities are particularly well-suited for such applications due to their high spectral resolution, narrow optical linewidths \cite{Hunger_highfinesse}, and in-situ tunability \cite{Saavedra:21}. Recent work has demonstrated the integration of G4V centers into diamond membranes, enabling coherent optical control and cavity coupling \cite{zifkin_lifetime_2024, herrmann_coherent_2024, rugar_quantum_2021, PhysRevApplied.23.034050}. However, the fabrication of high-quality, ultra-thin diamond membranes remains technically challenging. Surface roughness and interface imperfections can significantly degrade the quality of the coupled system. A scalable and practical alternative is to host color centers in nanodiamonds (NDs) \cite{PhysRevApplied.21.054032, fehler_hybrid_2021, greentree_nanodiamonds_2016, FP_NV_RT}. NDs offer several advantages: size-selective fabrication minimizes scattering losses, pre-characterization of individual emitters and deterministic integration into the cavity, eliminating the need for lateral cavity scanning.

Here, we demonstrate controlled coupling of a single \SnV center hosted in a ND by coupling its ZPL to a fully tunable FP microcavity. The proven single-photon nature of the \SnV center under off-resonant excitation allows direct detection of coherent photons from the ZPL. The narrowing of the single \SnV optical transitions at cryogenic temperatures allows the operation of the ND-cavity system in the weak coupling regime \cite{fox_optical_2012}. Further cooling to 4 K leads to an additional reduction of the linewidth transitioning the operation of the system from the bad-emitter (\SI{100}{K}) to the bad-cavity regime. In the latter regime, we observe a reduction in the excited-state lifetime corresponding to a Purcell factor above 1.7, confirming cavity-enhanced emission.

\section{\label{sec:citeref}Assembly of the hybrid ND–Cavity System}

\subsection{Experimental platform}

Our platform is a fully tunable hemispherical FP microcavity formed by a CO2 laser-ablated concave mirror and a macroscopic flat mirror. The measured radius of curvatures (ROC) of the spherical structure are $ROC_x \approx$ \SI{25}{\micro\meter} and $ROC_y \approx $ \SI{22}{\micro\meter} resulting in an ellipticity of 14 \%. The cavity consists of two distributed Bragg reflector (DBR) mirrors symmetrically designed.
The coating presents a first stopband centered at $\lambda$ = 620 nm with a transmission T = 650 ppm corresponding to a coating-limited finesse of $F_{ideal,618} = 4800 $. The second stopband in the green wavelength range, with a transmission of 6400 ppm, has a coating-limited finesse value of 400.

The cavity is embedded in a titanium bucket (purple rectangle in figure \ref{fig:Cavity_ND}a)). The flat cavity mirror is mounted on three nanopositioners (2xANPx101 and ANPz102, attocube systems AG) which enable precise tuning of the cavity length. A multimode fiber (MM), positioned below the cavity, collects the transmitted laser signal. A mode-matched achromatic lens couples the excitation laser into the cavity mode and collects, in reflection, the emitted fluorescence signal. To optimize coupling efficiency the achromatic lens is mounted on three nanopositioners allowing x,y and z tunability. The experimental setup including the microcavity, optical filters, and excitation/detection fibers is placed within a liquid helium bath cryostat (dashed rectangle).

\subsection{\texorpdfstring{Creation and positioning of the single \SnV center}{Creation and positioning of the single SnV- center}}

To create \SnV centers in NDs, size-selected diamond particles with an average size of \SI{180}{nm} \cite{Pureon_NDs} are implanted with a broad beam of $^{120}Sn^-$ ions at an energy of 60 keV and a fluence of $ 1\: x\: 10^{11} \:ions/cm^{2}$ at the Torino 100 keV ion implanter (LiuTo) \cite{Implanter_Turin}. To promote the formation of color centers, the sample is annealed in high vacuum for one hour at 1200 $^\circ\text{C}$ and in oxygen at 405 $^\circ\text{C}$ for one hour \cite{Nds_Implantation}. 

\begin{figure*}[t!]
\centering
\includegraphics[width =\textwidth]{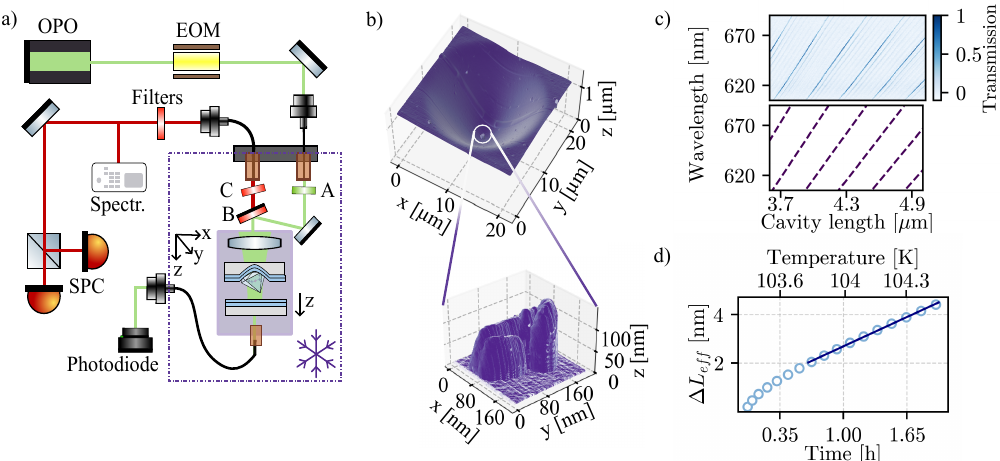}
\caption{ Assembled hybrid ND-cavity system. a) Schematic of the experimental setup. The single mode excitation fiber guides the laser signal from the optical parametric oscillator (OPO) and electric optical modulator (EOM) to the bath cryostat (purple-dashed box), which houses the microcavity, the achromatic lens (purple box), optical filters (A, C), the dichroic mirror (B), and the excitation and detection fibers. The detection fibers lead the signal to the spectrometer (Spectr.), the single photon counters (SPC) and photodiode. b) Atomic force microscope (AFM) scan of the spherical structure containing the ND. The measured ROCs are $ROC_x \approx$ \SI{25}{\micro\meter} and $ROC_y \approx $ \SI{22}{\micro\meter}. The zoom-in shows a high resolution AFM scan of the ND which size is approximately (\SI{150}{nm} x \SI{150}{nm} x \SI{80}{nm}). c) Upper panel: Cavity transmission spectra with the integrated ND probed by a white light diode (WLED) source showing the wavelength-dependent position of the cavity resonances at different cavity length. Lower panel: Theoretical distribution of the fundamental mode wavelengths for an empty cavity, calculated with equation \ref{eq:num}. d) Shift of the cavity length ($\Delta L_{eff}$) due to the rise in temperature measured over 2 hours. The dots represent the estimated values of $\Delta L_{eff}$, calculated from the free spectral range (FSR) determined using the WLED spectrum (see Appendix \ref{sec:cavity_para}). The blue line represents a linear fit with a slope of $\alpha = (5.1 \pm 0.03 ) \cdot 10^{-6} \; K^{-1}$.}
    \label{fig:Cavity_ND}
\end{figure*}

The sample is characterized off-resonantly with a 532 nm laser at room temperature (RT) using a custom-built confocal microscope. Once a single emitter with favorable optical properties is identified, it is transferred approximately to the center of the hemispherical cavity structure using an atomic force microscope (AFM) \cite{fehler_hybrid_2021, feuchtmayr_enhanced_2023}. An AFM scan of the spherical structure confirms the success of the transfer (figure \ref{fig:Cavity_ND}b)). A higher-resolution AFM image (zoom-in) reveals the shape and the size of the ND, approximately (\SI{150}{nm} x \SI{150}{nm} x \SI{80}{nm}). 
A room-temperature optical characterization of the same \SnV center confirms that the transfer process preserves the emitter optical properties. In table \ref{tab:free_space} we summarize the values of the ZPL, lifetime ($\tau_0$) and the saturation parameters ($P_{sat}$, $I_{sat}$) from the free-space characterization, see Appendix \ref{free_space}.

\begin{table}[ht]
    \centering
    \begin{minipage}{0.45\textwidth} 
    \begin{tabular}{l| r| l} 
       \textbf{Parameter}   & \textbf{Value} & \textbf{Units} \\
         \hline
        $ZPL$ & $620.1 \pm 0.1$ & $nm$ \\
        $FWHM$ & $6.20 \pm 0.03$ & $nm$ \\ 
         $g^2(0)$ & $0.25 \pm 0.04$ & \\
         $Lifetime$ $g^2(t)$ & $24 \pm 2 $& $ns$\\
         $Pulsed$ $lifetime$ & $21.7 \pm 0.3$ & $ns$\\
         $P_{sat}$ &  $1.20 \pm 0.08$ & $mW$ \\
         $I_{sat} $ & $185 \pm 6$ & $kC/s$ \\
       \end{tabular}
      \end{minipage}
    \caption{Free-space optical properties of the \SnV center at RT, under off resonant excitation, after the transfer into the concave cavity mirror.}
    \label{tab:free_space}
\end{table}

\subsection{Characterization of the hybrid ND-cavity system}\label{parIII}

After transferring the ND into the hemispherical structure, we proceed with the characterization of the resulting ND-cavity system. As a first step, we perform multiple cavity transmission scans. By evaluating the ratio of the free spectral range (FSR) to the full width at half maximum (FWHM) of the transmission resonances, we extract an average finesse value of $F_{exp,618} = 4600 \pm 500$ at 618.7 nm. This value is in good agreement with the ideal finesse $F_{ideal,618}$, indicating that the presence of the ND does not introduce scattering losses.

The cavity dispersion is characterized by transmission white-light diode (WLED) spectra as a function of the cavity length (upper panel in figure~\ref{fig:Cavity_ND}c)). The good agreement between experimental data and the simulation of the fundamental mode wavelengths (lower panel), based on the model for an empty cavity, 

\begin{equation}
    \nu_m = \frac{c}{2L}\left(m + \frac{cos^{-1}\left(\sqrt{1-L/ROC}\right)}{\pi}\right)
    \label{eq:num}
\end{equation}

confirms that the inclusion of the ND at such short cavity lengths does not significantly distort the mode dispersion (see figure \ref{fig:mode_dispersion} in Appendix \ref{mode_disp} ). 

Subsequently, the setup is indirectly cooled by a surrounding liquid nitrogen chamber. At approximately \SI{100}{K}, the system reaches thermal equilibrium, providing relative passive thermal stability for the cavity. To monitor stability, transmission WLED spectra are acquired continuously over a 2-hour period (see figure \ref{fig:cavity} c) Appendix \ref{sec:cavity_para}). During this time, a drift in the effective cavity length ($L_{eff}$) of approximately 5 nm is observed. The drift is attributed to a gradual temperature increase over the course of the measurement. To confirm this hypothesis, the drift in cavity length is fitted as a function of temperature using a linear model (figure~\ref{fig:Cavity_ND}d)). The resulting slope corresponds to a thermal expansion coefficient (CTE) of $\alpha = (5.10 \pm 0.03 ) \cdot 10^{-6} \; K^{-1}$. The good agreement of the estimated value with the reported CTE of titanium, $\alpha = 6.8 \cdot 10^{-6} \; K^{-1}$  \cite{king1938crystal} supports our interpretation.

\section{\texorpdfstring{Cavity coupling of the single SNV$^-$ center}{Cavity coupling of the single SnV- center}}

Off-resonant measurements are performed by exciting and detecting through the fundamental modes of the cavity. To enable simultaneous excitation and detection of the ZPL emission, the resonance condition $L_{eff} = m \cdot\frac{\lambda_m}{2}$ must be satisfied for both excitation ($\approx$ \SI{533.3}{nm}) and detection ($\approx$ \SI{618.5}{nm}) wavelength.

\begin{figure*}[t!]
    \includegraphics[width =\textwidth]{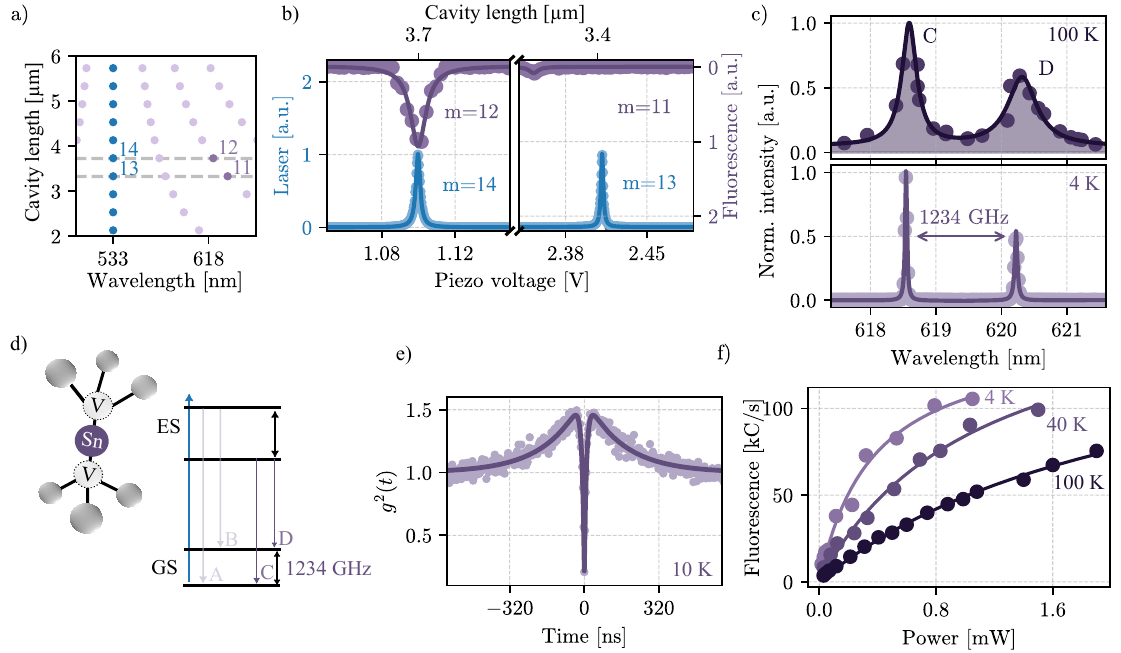}
    \caption{ Characterization of the single \SnV center through the cavity mode at \SI{10}{K} and \SI{100}{K}. a) Simulated resonance condition for FP cavity for different cavity length assuming a fixed excitation wavelength of \SI{533.3}{nm}. Laser resonances as blue dots and corresponding detection resonances as purple dots. The effective cavity lengths used in plot b) are indicated by the horizontal dashed lines (gray). b) Cavity length scans over two cavity resonances at \SI{4}{K} with a fixed laser wavelength \SI{533.3}{nm}. Transmitted laser signal in blue and reflected cavity coupled fluorescence signal in purple. Fluorescence signal is detected only when exciting through mode number $m_{exc}$ = 14. When $m_{exc}$ = 13 the emitter is coupling to a higher-order mode instead. c) PL spectrum of the single \SnV center revealing the fine structure (at low temperature). C and D transitions are fitted with a Lorentzian curve both at \SI{100}{K} and \SI{4}{K}, resulting in a ZPL position of $(618.5\:\pm\:0.1)$ nm and $(620.3 \: \pm \:0.1)$ nm respectively. d) On the left: atomic structure of \SnV centers in the diamond lattice. The tin atom (Sn) lies in between two adjacent carbon atoms (V). On the right: energy level scheme of the measured \SnV center, showing a ground state splitting of \SI{1234}{GHz}. The blue line represents the laser excitation. e) Second order autocorrelation measurement at \SI{63}{\micro\watt} showing antibunching with single-photon purity, without background correction: $g^2(0) = 0.21 \pm 0.03$. f) Saturation curves at 10, 40 and \SI{100}{K}. The data points represent the average photon counts acquired over a 30-second trace, while the solid line corresponds to the fitted function.}
    \label{fig:coupling}
\end{figure*}

A simulation of the resonance condition, performed with the excitation wavelength fixed, reveals that for $L_{eff} \approx$ \SI{3.7}{ \micro\meter} the second next resonance wavelength matches the ZPL of the \SnV center (figure \ref{fig:coupling}a)). This corresponds to excitation and detection through the mode numbers $m_{exc} = 14$ and $m_{det} = 12$. For mode numbers higher or lower than $m_{exc} = 14$, the ZPL is spectrally filtered out by the cavity and no significant counts are detected.

To verify the simulation, we experimentally tune the cavity length across two excitation resonances while simultaneously recording the laser transmission signal (blue curve, figure \ref{fig:coupling}b)) and the fluorescence signal in reflection (purple curve). Pronounced fluorescence is observed only when $m_{exc} = 14$ and $m_{det} = 12$, confirming cavity-emitter coupling at $L_{eff} \approx$ \SI{3.7}{ \micro\meter}. A weak fluorescence signal is also detected for $m_{det} = 11$, suggesting partial coupling to a higher-order cavity mode.

The fluorescence resonance appears broadened compared to the laser transmission due to the lower piezo-scanning speed, which is necessary to increase sensitivity to the lower photon counts. The extended scan time increases sensitivity to mechanical and environmental drifts, leading to additional broadening.

%All subsequent measurements are performed using this resonant configuration, with slight adjustments to the cavity length if needed. 
PL spectral probing at \SI{100}{K} is shown in figure~\ref{fig:coupling}c) (upper panel). The excitation laser, an optical parametric oscillator (OPO) (HÜBNER Photonics C-WAVE GTR), is tuned together with the cavity length. For each excitation wavelength, a PL spectrum is recorded (see Appedix \ref{spec_100K}). By summing all spectra, two distinct optical transitions, labeled C and D, become visible. These transitions arise from the characteristic splitting of orbital levels due to spin–orbit interaction, which originates from the $D_{3D}$ symmetry of the \SnV center in diamond  \cite{Becher_2014}. This symmetry results from the tin atom occupying a split-vacancy configuration (left panel, figure ~\ref{fig:coupling}d)).
Spin–orbit coupling leads to the formation of four orbital levels: two from the ground state (GS) splitting and two from the excited state (ES) splitting \cite{harris_hyperfine_2023}. The corresponding optically allowed transitions are labeled A, B, C, and D in figure \ref{fig:coupling}d) (right panel). At \SI{4}{K} only transitions C and D occur and, only at higher temperatures ($>$\SI{70}{K}) transitions A and B become visible \cite{gorlitz_2020}. In our experiment, due to the weak emission of transitions A and B we can only observe transitions C and D also at higher temperatures. At \SI{100}{K}, these transitions are fitted with a Lorentzian, and the resulting wavelengths are $(618.6\: \pm \:0.1)$ nm and $(620.3 \: \pm \:0.1)$ nm, respectively. The respective linewidths are (210 $\pm$ 20) GHz and (410 $\pm$ 20) GHz. 

The PL spectrum at \SI{4}{K} reveals a narrowing of the optical transitions to the spectrometer resolution limit of 16 GHz. The emission wavelengths at this temperature are $(618.54\: \pm \:0.03)$ nm and $(620.22 \: \pm \:0.03)$ nm, respectively. The resulted GS splitting is 1234 GHz (lower panel figure \ref{fig:coupling}c)). The GS splitting is significantly larger than the theoretical predicted value of 850 GHz \cite{Gali_2018}, suggesting substantial strain in the surrounding ND lattice favorable for high-fidelity microwave spin control \cite{Brevoord2025, PhysRevX.13.031022}.

Single photon emission is confirmed by Hanbury Brown–Twiss measurements. The emitter is excited with the OPO at \SI{533.25}{nm}, enabling detection on resonance at \SI{618.54}{nm}. The collected fluorescence is split by a 50:50 beam splitter and directed to two SPCs. To account for the prominent bunching observed in figure \ref{fig:coupling}e) the data points are fitted with a three-level model equation \cite{Steck, neu_single_2011}.A minimum value of $g^{2}(0) = 0.21 \pm 0.03$, without any background correction, demonstrates the single-photon purity of the \SnV center. The single-emitter behavior under off-resonant excitation enables direct detection of ZPL photons. %Thus, all subsequent measurements don't rely on phonon sideband collection, highlighting the purity and optical stability of the C optical transition.

An investigation of the saturation behavior of the emitter at \SI{4}{K}, \SI{40}{K}, and \SI{100}{K} (figure \ref{fig:coupling}f)) reveals that as the temperature decreases, the saturation power also decreases (table \ref{tab:saturation}).

%The resulting values of saturation power ($P_{Sat}$) and saturation intensity ($I_{Sat}$), from the saturation fit \cite{fox_optical_2012}, are summarized in table .

\begin{table}[h!]
    \centering
    \begin{tabular}{c|c|c}
      Temperature [K] & $P_{\text{Sat}}$ [mW] & $I_{\text{Sat}}$ [kC/s] \\
      \hline
      $10$   & $0.37 \pm 0.06$ & $150 \pm 10$ \\
      $40$   & $1.1 \pm 0.2$   & $180 \pm 20$ \\
      $100$  & $2.2 \pm 0.2$   & $162 \pm 9$ \\
    \end{tabular}
    \caption{Saturation power and intensity values resulting from the fit $I = I_{\text{sat}} \cdot P / (P_{\text{sat}} + P)$.}
    \label{tab:saturation}
\end{table}

This behavior is governed by the relationship $P_{sat} \propto 1/\sigma$
where $\sigma$ is the absorption cross-section of the emitter \cite{basov1963saturation}. At lower temperatures, the optical linewidth of the emitter narrows due to reduced phonon interactions and spectral diffusion, effectively increasing $\sigma$. The higher absorption cross-section enables more efficient photon absorption, reducing the optical power necessary to reach saturation.

\begin{figure*}[t!]
    \includegraphics[width =\textwidth]{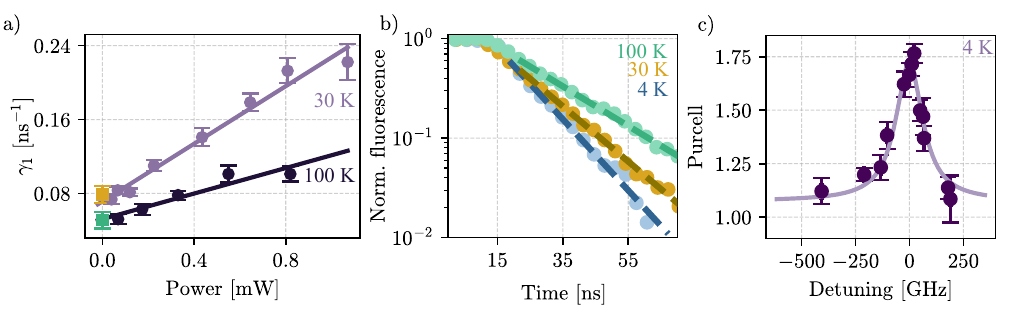}
    \caption{Cavity-induced enhancement of the spontaneous decay rate across different temperatures. a) Decay rate ($\gamma_1$), extracted from fitting $g^2(t)$ functions measured under varying excitation powers. The solid line represents the linear fit. The estimated lifetime values are $(14 \pm 1)$ ns at \SI{30}{K} and $(19 \pm 2)$ ns at \SI{100}{K}. The green and yellow square are the lifetime values estimated from the pulse measurements. b) Logarithmic scale of pulsed lifetime measurements at three different temperatures (\SI{4}{K}, \SI{40}{K} and \SI{100}{K}). The dashed lines are the linear fits to extrapolate the lifetime values. The obtained lifetime values are $(12.2 \pm 0.3 )$ ns at \SI{4}{K}, ($15.8 \pm 0.3$) ns at \SI{40}{K} and $(21 \pm 1)$ ns at \SI{100}{K}. c) Purcell factor, estimated as the ratio
    in between free-space lifetime and cavity modulated one as a function of cavity detuning. The cavity length varies between fully off- and on- resonance. }
    \label{fig:detuning}
\end{figure*}

\section{\texorpdfstring{Cavity-enhanced interaction of the SNV$^-$ center in the weak coupling regime}{Cavity-enhance interaction of the SnV- center in the weak coupling regime}}\label{par:lifetime}

In the bad-cavity regime the lifetime of the emitter is reduced due to the Purcell effect. The overall Purcell enhancement can be estimated as \cite{Janitz:20}: 
%This change in lifetime can be used to estimate the Purcell factor $F_p$, and, in turn, to determine the coupling strength between the emitter and the cavity mode. 

\begin{equation}
    F_p = \frac{\tau_o}{\tau_p} 
    \label{eq:Purcell}
\end{equation}

where $\tau_p$ and $\tau_0$ (\ref{tab:summary}) are the cavity-modified and the free-space spontaneous emission lifetime, respectively. 

To estimate $\tau_p$ we measure power-dependent second-order autocorrelation functions ($g^2(t)$) and fit the data with the three level model equation \cite{Steck}. Since the decay rate ($\gamma_1$) varies linearly with the excitation power the resulting values are fitted with the model $ \gamma_1 = m \cdot P + 1/\tau$ \cite{berthel_photophysics_2015}.

Lifetimes of $(14 \pm 1)$ ns and $(19 \pm 2)$ ns are extracted at zero excitation power at $\sim$\SI{30}{K} and \SI{100}{K}, respectively (figure \ref{fig:detuning}a)). The measurements are performed with the cavity maintained on resonance with the emitter frequency by actively adjusting the cavity length to compensate for temperature-induced drifts. The resulting lifetime values are confirmed through pulsed lifetime measurements. The OPO laser, fixed at 533.25 nm, is pulsed off with an electric optical modulator (EOM) while recording the time delay between the excitation pulse and the detected fluorescence photons. By fitting the data with an exponential decay we obtain the following lifetime values: $(12.2 \pm 0.3 )$ ns at \SI{4}{K}, ($15.8 \pm 0.3$) ns at \SI{40}{K}, and $(21 \pm 1)$ ns at \SI{100}{K} (figure \ref{fig:detuning}b)). As shown in figure \ref{fig:detuning}a), the pulsed lifetime measurements (yellow and green square) intersect the linear extrapolation of the power-dependent $\gamma_1$ values at zero excitation power. The agreement confirms the reliability of the extracted lifetime values. The lifetime reduction indicates that the system operates in the weak coupling regime.

At \SI{100}{K}, the emitter linewidth ($\gamma = $ \SI{210}{GHz}) exceeds the cavity-field decay rate ($\kappa=$\SI{15}{GHz}), placing the system in bad-emitter regime ($\gamma > \kappa,g$) \cite{PhysRevLett.110.243602}. In this regime, only a fraction ($\frac{\kappa}{\gamma}$) of the spontaneous emission couples to the cavity mode \cite{PhysRevA.88.053812}. As a result, spontaneous emission is funneled into the cavity mode without significantly modifying the emitter’s radiative lifetime \cite{PhysRevLett.114.193601, PhysRevApplied.13.064016, hausler_tunable_2021}.  This explains the weak observed Purcell enhancement: the measured lifetime $(21 \pm 1)$ ns is only marginally shorter than the free-space value (see table \ref{tab:free_space}).

In contrast, at lower temperatures, the narrowing of the emitter optical transitions enables operation in the bad-cavity regime ($\gamma  \; < \; \kappa$), where the cavity-field decay rate dominates. Under these conditions, we estimate a Purcell factor of $F_p = 1.78 \pm 0.04 $. The definition of $F_p$ (equation \ref{eq:Purcell}) includes the emission into the incoherent phonon sideband and the free-space continuum\cite{yurgens_cavity-assisted_2024}. However, since only the fraction of emission coupled into the C optical transition is enhanced by the cavity, we correct $F_p$ with a factor $\epsilon = 0.36$. The correction factor $\epsilon$ accounts for non-unity quantum efficiency (approximately 80\% \cite{Iwasaki2017}), a limited Debye–Waller factor ($\sim$56\%, estimated from the free-space photoluminescence spectrum), and off-resonant branching ratio into C transition (80 \% \cite{rugar_quantum_2021}). Accounting for all these effects, we estimate a corrected Purcell enhancement of $F_{p,ZPL} = 4.9$.

The Purcell factor ($F_{p}$) follows a Lorentzian dependence on cavity detuning \cite{englund_controlling_2005, rugar_quantum_2021, Chen:24}:

\begin{equation}
   F_{p} = F_{cav} \left(\frac{\vec{E}(r) \vec{\mu}}{|\vec{E}_{max}||\vec{\mu}|}\right)^2 \frac{1}{1+ \left(2Q\left(\frac{\lambda}{\lambda_{cav}} - 1\right)\right)^2} + F_{fp}.
   \label{eq:detuning}
\end{equation}

The first term accounts for spatial and spectral mismatch between the emitter dipole $\vec{\mu}$ to the cavity field $\vec{E}$, while $F_{fp}$, accounts for background modifications to the emission rate due to the cavity environment. For perfect alignment, the enhancement is given by $F_{cav}$ + $F_{fp}$, where the theoretical Purcell factor $F_{cav}$ is \cite{Janitz:20}:

\begin{equation}
      F_{cav} = \frac{3}{4\pi^2} \cdot \left(\frac{\lambda_{c}}{n}\right)^3 \cdot \frac{Q}{V_{mode}}
\label{eq:purcell_theo}
\end{equation}

For our FP cavity, operating at a wavelength of $\lambda_c$ = 618.5 nm, an estimated quality factor of $Q_{ideal}$ = 56400 and mode volume of $V_{mode} = 21 \lambda_c^3$, we obtain $F_{cav} = 205$. However, since the ND is located on the spherical structure rather than at the field maximum of the cavity mode, it experiences weaker coupling. 
To quantify the effect of this spatial mismatch, we use the ratio $(\frac{\omega_0}{\omega\left(L\right)})^2$, which effectively describes the reduction of the optimal coupling due to the position \cite{bayer_optical_2023}. Incorporating this spatial misalignment, we obtain a corrected Purcell factor of $F_{cav} = 173$.

To understand the significant discrepancy between the theoretically predicted Purcell factor and the experimentally measured value, we perform a detailed analysis of the system’s spectral properties by measuring the Purcell enhancement as a function of cavity-emitter detuning (figure ~\ref{fig:detuning}c)). The detuning-dependent lifetime measurements allow us to extract the effective cavity-field decay rate and assess whether additional broadening mechanisms are degrading the cavity performance. Cavity detuning is achieved by tuning the excitation wavelength and adjusting the cavity length to maintain resonance conditions for $m_{det}$ = 12 and $m_{exc}$ = 14. For every cavity detuning, we perform pulsed lifetime measurements, with increasing acquisition time for larger detuning to account for minor fluorescence counts. 

Fitting the detuning-dependent data yields an effective cavity-field decay rate of $\kappa_{\text{exp}} = (160 \pm 30)$ GHz, nearly an order of magnitude larger than the expected value of 15 GHz. This broadening arises from fast mechanical vibrations and thermal drift, which shift the cavity resonance during acquisition \cite{ruf_resonant_2021}. Despite manually compensating for slow drifts, rapid fluctuations reduce the effective quality factor to $Q_{exp} = (3100 \pm 600)$. Under these realistic conditions, the maximum achievable Purcell factor is limited to $F_{\text{vib}} = 10 \pm 2$. This value reflects the upper bound imposed by the degraded Q factor.

Comparing $F_{\text{vib}}$ with the measured Purcell factor $F_p = 4.9$ (formula \ref{eq:Purcell}) provides insight into the efficiency of coupling between the emitter and the cavity mode. We obtain an effective spatial alignment factor of approximately 49\%. This value corresponds to the fraction of the C emitter dipole moment $\mu$ that is aligned with the cavity electric field (E), assuming perfect spatial overlap and including. Conversely, when we do not explicitly correct for the branching ratio, we find an alignment factor of about 38\%. This lower value implicitly includes both the dipole misalignment and the reduction in effective coupling strength due to the branching ratio. Thus, improving spatial alignment of the emitter within the cavity field, by moving or rotating the ND, could lead to a higher Purcell factor and further lifetime reduction.

\section{Conclusion}

We demonstrate the successful coupling of a single \SnV center in a ND to a FP cavity. The small size of the ND minimizes scattering losses \cite{PhysRevLett.110.243602, bayer_optical_2023, feuchtmayr_enhanced_2023}. As a result, the cavity maintains a finesse of $4600 \pm 500$, comparable to that of a bare cavity. Furthermore, we show that the dispersion of the cavity modes remains unaffected by the presence of the ND and can be accurately predicted using the model for an empty cavity. The presence of a single \SnV center in the ND, as confirmed by the observed minimum of the autocorrelation function $g^2(0) = 0.21 \pm 0.03$, enables the detection of coherent photons from the ZPL. This coherence is crucial for the generation and establishment of indistinguishable photons, a key requirement for scalable quantum networks and spin–photon interfaces \cite{Becher_chargeCicle, PhysRevB.109.085414, Li2015}. By cooling the system to cryogenic temperatures, we operate the system in the weak coupling regime. Specifically, at \SI{100}{K}, a linewidth of 210 GHz allows operation in the bad-emitter regime, where we observe single photon emission into the cavity mode over a very broad spectral range, which can be exploited to realize a continuously tunable, narrow-band single photon source \cite{riedel2017deterministic, FP_NV_RT, PhysRevApplied.13.064016,PhysRevLett.110.243602}. Further cooling narrows the optical transition C and shifts the system into the bad-cavity regime, where a clear lifetime reduction, corresponding to a Purcell factor of $1.78 \pm 0.04$ is observed. This enhancement is crucial for realizing highly efficient photon sources, as it increases the probability of photon emission into the desired cavity mode, improving brightness and emission rate for quantum photonic applications. Probing of the fine structure, reveals an increased GS splitting of 1234 GHz favorable for high-fidelity microwave spin control \cite{Brevoord2025, PhysRevX.13.031022}. 

By detuning the cavity from fully on resonance to off-resonance, we probe the emitter’s lifetime under varying coupling conditions. These measurements reveal a coupling efficiency of approximately 49\%, indicating that precise manipulation of the nanodiamond (ND) position using an AFM could further enhance the interaction between the cavity and the C optical transition \cite{lettner_controlling_2024, fehler_hybrid_2021}. Moreover, we observe that rapid mechanical vibrations of the cavity degrade its performance, limiting the maximum achievable Purcell factor by approximately an order of magnitude. Therefore, improving the mechanical stability of the cavity is crucial for further increasing both the Purcell enhancement and emitter–cavity coupling efficiency \cite{pallmann_highly_2023}.

These results demonstrate that the integration of single \SnV centers in size-controlled NDs within tunable FP microcavities provides a practical route toward versatile and tunable quantum light–matter interfaces.

\section*{Acknowledgment}
This project was funded by the German Federal Ministry of Research, Technology and Space (BMRTS) under the project QR.N (16KIS2208). We acknowledge additional support from the LasIonDef project (LASer fabrication and ION implantation of DEFects as quantum emitters), funded by the European Research Council under the Marie Skłodowska-Curie Innovative Training Networks program. Further funding was provided by the European Union’s QuantERA Program through the SensExtreme project (499192368), which is gratefully acknowledged.

The author thanks Prof. Dr. Christine Kranz, Dr. Gregor Neusser, and the Focused Ion Beam Center at Ulm University for their support with FIB milling. Jens Fuhrmann is acknowledged for the annealing of the samples, and Nadia Elia for her assistance during the measurements. We are grateful to Pureon AG for providing the nanodiamond powder, and to the service team at HÜBNER Photonics for their valuable support in optimizing the C-WAVE GTR widely tunable continuous-wave laser system.

\appendix
\section{Free-space characterization} \label{free_space}

Prior to assembling the Fabry–Perot (FP) microcavity, we characterize the optical properties of the single \SnV center at room temperature (RT), both before and after its transfer into the hemispherical structure of the cavity mirror. As no significant differences are observed between the two measurements, we report only the data obtained after the transfer. 

\begin{figure}[ht]
\includegraphics[]{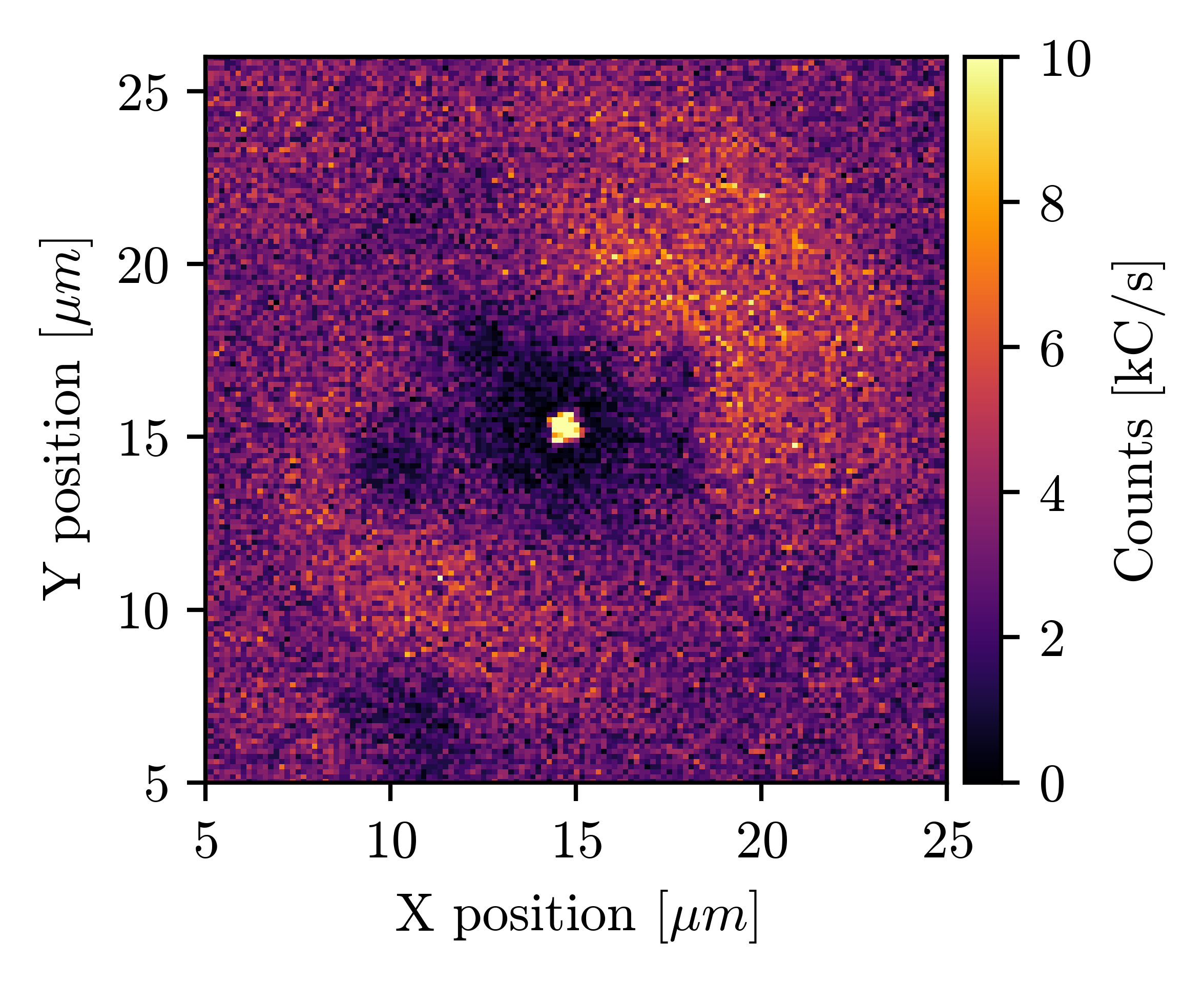}
    \caption{Confocal microscope scan of the nanodiamond inside the hemispherical structure.}
    \label{fig:confocal}
\end{figure}

\begin{figure*}[ht]
    \includegraphics[]{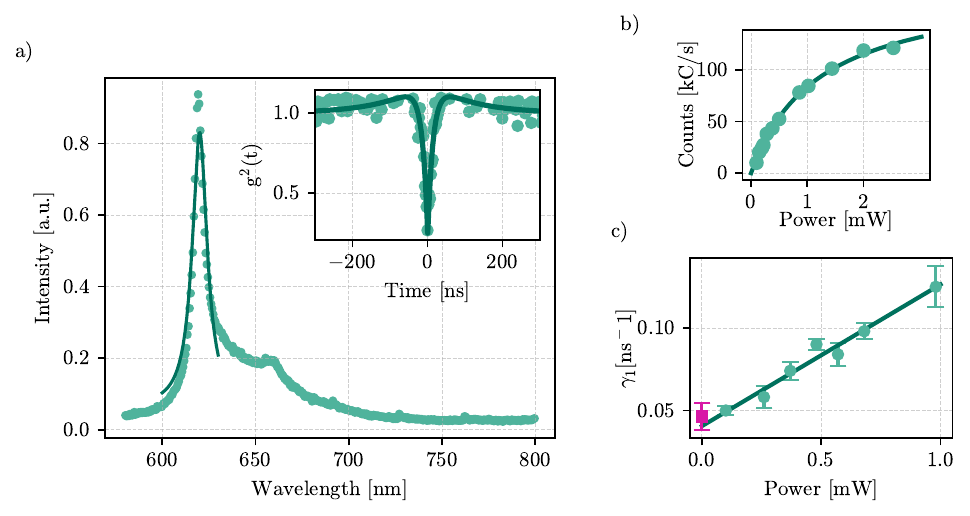}
    \caption{Room temperature characterization of the \SnV center inside the hemispherical structure. a) Off-resonant PL spectrum, the Lorentzian fit (solid line) results in a ZPL of ($619.7 \pm 0.2$) nm and a FWHM of 6 nm. Inset: second-order autocorrelation function proving the quantum nature of the emitter, $g^2(0) = 0.25 \pm 0.04$. b) Saturation curve, the data points are the average of the time trace recorded over \SI{30}{s}. The theoretical fit leads to a $I_{sat} = (185 \pm 6)$ kC/s and $P_{sat} =  (1.20 \pm 0.08 ) $ mW. c) Decay rate ($\gamma_1$), extracted from fitting the power dependent $g^2(t)$ curves, as a function of the excitation power. The solid line represents the linear fit leading to a lifetime of $(24 \pm 2)$ ns. The pink square is the lifetime estimated with the pulse measurements $\tau_{pulsed}$ = (21.7 $\pm$ 0.3) ns.
    } \label{fig:FR}
\end{figure*}

The characterization is performed using a home-built confocal microscope. Figure \ref{fig:confocal} shows the confocal scan under continuous off-resonant excitation at \SI{532}{nm}, where the nanodiamond (ND) appears as a bright spot approximately in the middle of the structure. After identifying the ND, we acquire the photoluminescence (PL) spectrum, shown in Figure \ref{fig:FR}a). The zero-phonon line (ZPL) is clearly visible and is well fitted by a Lorentzian function, yielding a center wavelength of $(620.1 \pm 0.1)$nm and a full width at half maximum (FWHM) of \SI{6}{nm}. The measured ZPL position is in good agreement with the typical ZPL reported for \SnV centers in bulk diamond \cite{Iwasaki2017, Rugar2020, Debroux_2021}. Additionally, a weaker peak is observed at \SI{660}{nm}, which is attributed to the phonon sideband, consistent with previous reports \cite{Iwasaki2017}. From the ratio of the integrated intensities of the ZPL and the phonon sideband, we estimate a Debye–Waller factor of approximately 56\%, in agreement with previously reported values \cite{gorlitz_2020, Iwasaki2017}.

The quantum nature of the emitter is confirmed with Hanbury Brown–Twiss measurements. To account for the prominent antibunching observed in the inset of figure \ref{fig:FR}a) the data points are fitted with a three-level model equation:

\begin{equation}
g^{2}(t) = 1 + c \cdot (\beta \cdot e^{- \gamma_1|t - t_0|} + (\beta -1)\cdot  e^{-\gamma_2|t - t_0|})
\label{eq:g2}
\end{equation}

where $\gamma_1$ and $\gamma_2$ are the decay rates of the excited state and shelving state, respectively. The other parameters are: $t_0$ the temporal offset, c determines how pronounced the antibunching is and $\beta$ weight the two exponential factors. The observation of antibunching with a minimum value of $g^2(0)$ = $0.25\pm0.04$, without any background correction, demonstrates the quantum nature and single-photon purity of the \SnV center. 

We further investigate the saturation behavior of the emitter (figure \ref{fig:FR}b)). Each data point corresponds to the time-averaged count rate over \SI{30}{s}. The resulting value of saturation power ($P_{Sat}$) and saturation intensity ($I_{Sat}$) are: $P_{Sat} = (1.20 \pm 0.08) \: mW$ and $I_{sat} = (185 \pm 6) \: kC/s$. 

\begin{figure*}[ht]
\includegraphics[]{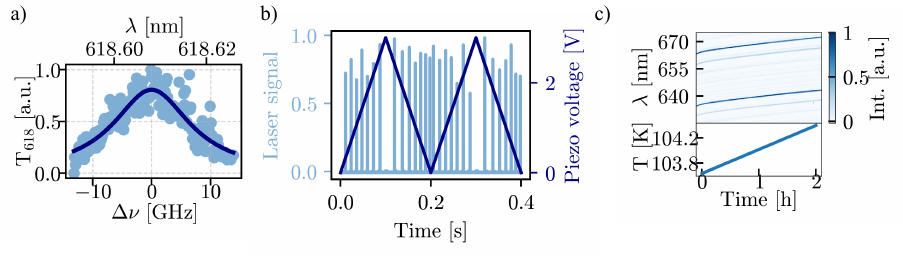}
    \caption{Cavity characterization at \SI{618.7}{nm}. a) Transmitted laser scan over a cavity resonance. The Lorentzian fit yields a cavity-field decay rate of $ \kappa = (15 \: \pm \: 1)$ GHz. b) Voltage ramp over \SI{100}{ms} (blue lines) while the transmitted laser signal is recorded with a photodiode. The resulting average finesse value is $F_{exp,618} = 4600 \pm 500$. c) Upper panel: 7200 transmitted white light diode spectra acquired over 2 hours. Lower panel: recorded temperature ([K]) during the spectra acquisition.}
    \label{fig:cavity}
\end{figure*}

To estimate the spontaneous decay rate in free-space at RT, we measure power-dependent second-order autocorrelation functions. The decay rate ($\gamma_1$) varies linearly with the excitation power \cite{khalid_lifetime_2015}, so the extracted values are fitted with the model $ \gamma_1 = m \cdot P + 1/\tau$ (figure \ref{fig:FR}c)). From the fit, we determine a lifetime of $\tau_0 = (24 \pm 2)$ ns at 0 excitation power.

To validate the result, we perform pulsed lifetime measurement by exciting the \SnV center with a pulsed 532 nm laser and recording the time delay between the excitation pulse and the detected fluorescence photon. The resulting exponential decay is fitted with a single exponential function. The resulting lifetime value is $\tau_{pulsed}$ = (21.7 $\pm$ 0.3) ns (pink square). 

The predicted and previously observed lifetimes for \SnV centers in bulk diamond range from 5 to 7 ns \cite{Gali_2018, PhysRevLett.124.023602, Iwasaki2017, tchernij_single-photon-emitting_2017, Iwasaki2017}. A recent study on single \SnV centers in NDs reveals a shorter lifetime ($\tau <$ \SI{4}{ns}) \cite{Fujiwara2025}. In contrast, our measured value is approximately four times longer. 

Several factors influence the radiative decay rate, including the local photonic density of states (LDOS) \cite{zahedian_modeling_2023, inam_tracking_2013, PhysRevLett.128.153602}, damage induced during sample preparation \cite{gorlitz_2020}, and the presence of non-radiative decay channels \cite{eremchev_detection_2023, cheng_laser_2024}.

The LDOS is sensitive to the emitter’s position relative to the diamond surface and the size of the ND. Specifically, a higher LDOS means more available states for the color center to transition into, leading to a shorter lifetime. Conversely, a lower LDOS can extend the lifetime \cite{zahedian_modeling_2023, PhysRevLett.128.153602}. This phenomenon has been observed in NV centers, where a correlation between ND size and emitter lifetime was reported \cite{inam_tracking_2013}. Conversely, for G4V centers in bulk diamond, for emitters implanted at depths less than 20 nm, shorter lifetimes have been observed, likely due to a more rapid non-radiative decay \cite{cheng_laser_2024} or delayed spontaneous emission dynamics \cite{eremchev_detection_2023}. Moreover, lattice damages induced by annealing and implantation processes influence spontaneous lifetime emission \cite{gorlitz_2020}.

In our case, it is plausible that the implantation process, especially considering the large size of the Sn atom, has introduced considerable lattice damage. Additionally, our $g^{(2)}(t)$ measurements reveal a non-radiative decay component with a long characteristic lifetime of approximately \SI{200}{ns}. Both lattice damages and the presence of a long-lived non-radiative channel contribute to extend the lifetime of the color center \cite{gorlitz_2020, cheng_laser_2024, inam_tracking_2013}. Although, the precise origin of the untypical long total lifetime remains unclear due to unknown defect depth and lattice conditions.

\section{Characterization of the experimental platform \label{sec:cavity_para}}

Figure \ref{fig:cavity}a) shows a cavity transmission measurement performed to determine the cavity-field decay rate. A resonant laser at \SI{618.61}{nm} is scanned twice over the cavity resonance, while the transmitted laser is detected with a single photon counter (SPC). A Lorentzian fit to the transmission peak yields a cavity-field decay rate of $(15 \: \pm \: 1)$ GHz. 

The finesse is determined by cavity scan length (figure \ref{fig:cavity}b)), where a voltage is applied to the z-piezo while recording the transmitted laser intensity with a photodiode. The piezo voltage is swept from \SI{0}{V} to \SI{3}{V} and back (dark blue trace), allowing multiple forward and backward piezo ramps to be analyzed to estimate an average finesse. Each resonance (light blue lines) is fitted with a Lorentzian. From the ratio of the free spectral range (FSR) to the full width at half maximum (FWHM), we extract an average finesse of $F_{exp,618} = 4600 \pm 500$ at \SI{618.7}{nm}.

\sloppy
Figure \ref{fig:cavity}c) displays the transmission spectra of a white light diode (WLED), recorded over a total duration of 2 hours, s 7200 showing individual spectra. To estimate the effective change in cavity length, $\Delta L_\text{eff}$, we track the shift in wavelength position of the fundamental cavity mode over time. Specifically, the shift in resonance wavelength between the spectra acquired at $t =0$ and t=\SI{2}{h} is used to quantify the cavity length variation, $\Delta L_\text{eff} =(\lambda_{res,2h} - \lambda_{res,0})/2$. The corresponding values of $\Delta L_\text{eff}$ are presented in Figure 1d) of the main manuscript.

\section{Cavity mode dispersion}\label{mode_disp}

\begin{figure}[ht]
\includegraphics[]{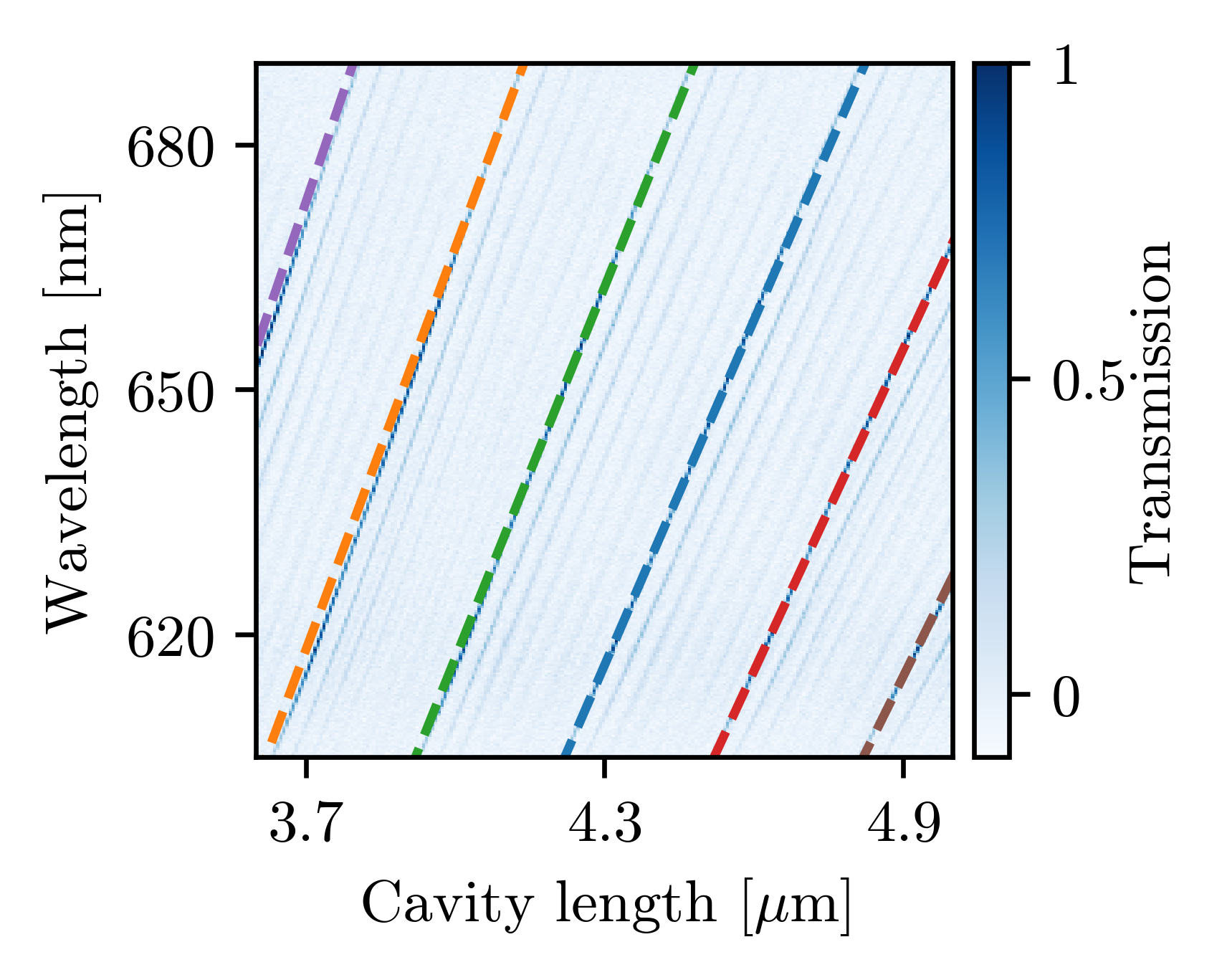}
    \caption{Cavity transmission spectra with the integrated ND probed by a white light diode (WLED) showing the wavelength-dependent position of the cavity resonances at different cavity lengths. Dashed lines are the simulated cavity resonances.}
    \label{fig:mode_dispersion}
\end{figure}

We further characterize the cavity mode dispersion to investigate the influence of the ND on the cavity field. Using a WLED, we record the wavelength-dependent cavity resonances as a function of the cavity length. The transmitted light is collected through the transmission fiber and recorded with a spectrometer. A total of 300 consecutive spectra are acquired while scanning the cavity length from \SI{5.0}{\micro\meter} to \SI{3.7}{\micro\meter}. The data reveal the presence of the fundamental mode along with higher-order modes, as shown in Figure~\ref{fig:mode_dispersion}. The dashed lines are the simulated data obtained with the model for empty cavity including the Gouy-phase \ref{eq:num}. 

For the simulation, an ROC of \SI{24}{\micro m} has been used and the longitudinal mode index m varies from 11 to 17.
The overlap of the simulated resonances and experimental data proves that introducing the ND in the hemispherical structure does not alter the mode profile.

\section{PL spectrum probing at 100 K}\label{spec_100K}

\begin{figure}[ht]
\includegraphics[]{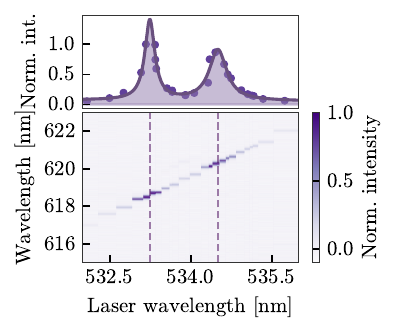}
    \caption{PL spectrum probing at \SI{100}{K}. Upper panel: Averaged counts from all the spectra. Lower panel: Cavity-modulated PL spectra at \SI{100}{K} as a function of the excitation laser wavelength. The dashed lines represent the laser wavelength used in the main paper to be on resonance with optical transitions C and D: (533.25 $\pm$ 0.03) nm or (534.50 $\pm$ 0.03) nm respectively.}
    \label{fig:PL100}
\end{figure}

PL spectral probing at \SI{100}{K} is performed by tuning both the excitation laser wavelength and the cavity length to enable detection of ZPL emission through $m_{det} = 12$ and green excitation through $m_{exc} = 14$. 

In figure \ref{fig:PL100} (lower panel) we report all the recorded spectra as a function of the excitation wavelength. The dashed vertical lines mark the position of the excitation laser obtained from the Lorentzian fit ($533.25 \pm 0.03$) nm and ($534.50 \pm 0.03$) nm for transition C and D respectively. The laser wavelength obtained for C transition is used in the main manuscript for measurements resonant with the emitter, without detuning.

\begin{figure*}[ht]
    \includegraphics[]{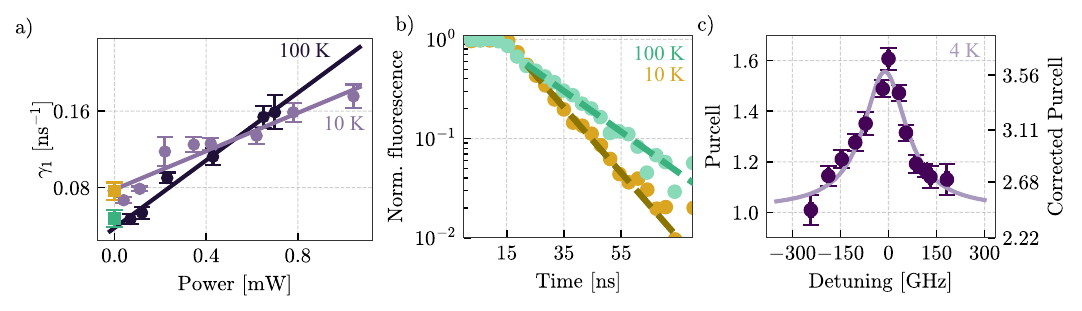}
    \caption{ Cavity-induced enhancement of the spontaneous decay rate of the optical transition D across different temperature. a) Decay rate ($\gamma_1$), extracted from fitting $g^2(t)$ functions measured under varying excitation powers. The solid line represents the linear fit. The estimated lifetime values are $(13 \pm 1)$ ns at \SI{30}{K} and $(21 \pm 6)$ ns at \SI{100}{K}. The green and yellow square are the lifetime values estimated from the pulse measurements. b) Logarithmic scale of pulsed lifetime measurements at three different temperatures (\SI{4}{K}, \SI{40}{K} and \SI{100}{K}). The dashed lines are the linear fits to extrapolate the lifetime values. The obtained lifetime values are $(13.1 \pm 0.3)$ ns at \SI{4}{K} and $(21\pm 2)$ ns at \SI{100}{K}. c) Purcell factor, estimated as the ratio in between free-space lifetime and cavity modulated one as a function of cavity detuning. The cavity length varies between fully off- and on- resonance. 
    } \label{fig:D_Purcell}
\end{figure*}

\section{Bad-emitter regime}\label{bad_emitter}

The total spontaneous decay rate ($\gamma$) of an emitter can be expressed as the sum of the emitter’s natural linewidth ($\gamma_0$), determined by its radiative lifetime, and the pure dephasing rate ($\gamma*$) arising from interactions with phonons, $\gamma = \gamma_0 + \gamma*$. At \SI{100}{K}, the phonon population in the host lattice is still substantial, leading to significant pure dephasing. In this regime, the dephasing rate dominates over the lifetime-limited contribution ($\gamma^* \gg \gamma_0$), and we approximate $\gamma \sim \gamma*$. Thus, we assume that the spontaneous decay rate is equal to the FWHM of the PL spectrum acquired at \SI{100}{K}. This corresponds to $\gamma^*_C = (210 \pm 20)$ GHz and $\gamma*_D = (410\pm20)$ GHz.

Due to the broad emitter linewidth ($\gamma \gg \kappa$) only a fraction $\frac{\kappa}{\gamma^*}$  of the total emission is spectrally resonant with the cavity mode and effectively couples into it \cite{PhysRevA.88.053812}. 

In this regime, the Purcell factor can be approximated as: \cite{PhysRevLett.110.243602}:

\begin{equation}
   F_{bad-emitter} \approx  \frac{4g^{2}}{\kappa + \gamma_0} \times \frac{\kappa}{\gamma^*}.
\end{equation}

Since in the bad-emitter regime $\gamma^* > g$ we expect a $F_{bad-emitter} < 1$. As a consequence, the cavity would not alter the emitter's lifetime, but rather channel the emission into a well-defined cavity mode \cite{PhysRevLett.110.243602}. With this model we can explain the observation of no lifetime reduction both optical transitions.

\section{Purcell enhancement of D optical transition}\label{Purcell_D}

As for optical transition C, we investigate the interaction between the cavity electric field and D optical transition. The Purcell enhancement is estimated with equation \ref{eq:Purcell}. The cavity-modulated lifetime ($\tau_p$) is estimated through power-dependent second-order autocorrelation functions ($g^2(t)$) and pulsed lifetime measurements. 

From the linear for of the obtained $\gamma_1$ values, lifetime values of $(13 \pm 1)$ ns and $(21 \pm 6)$ ns are observed at a temperature of $\sim$\SI{10}{K} and \SI{100}{K}, respectively (figure \ref{fig:D_Purcell}a)). 
The resulting lifetime values through pulsed lifetime measurements are $\tau_{10K,D} = (13.1 \pm 0.3 )$ ns at \SI{10}{K} and ( $\tau_{100K,D} = (21 \pm 2)$ ns at \SI{100}{K} (figure \ref{fig:D_Purcell}b)). The measured pulsed lifetime (yellow and green squares in figure \ref{fig:D_Purcell}a)) intersect with the linear extrapolation of the power-dependent values at zero excitation power, confirming the reliability of the extracted lifetimes.

At \SI{100}{K}, no lifetime reduction is observed, consistent with the operation of the cavity in the bad-emitter regime. 
In contrast at $\approx$\SI{10}{K}, we measure a lifetime reduction leading to a Purcell factor of $F_p = 1.67$. Further deviations from the ideal Purcell enhancement arise from non-unity quantum efficiency (approximately
80\% \cite{Iwasaki2017}) and a limited Debye–Waller factor ($\sim$56\%). Each of these factors reduces the fraction of emission coupled into the D optical transition. Accounting for all these effects, we estimate a corrected Purcell enhancement of F$_{p,ZPL}$ = 3.72.  
By detuning the cavity resonance from the \SnV center frequency and performing pulsed lifetime measurements (Figure \ref{fig:D_Purcell}c)), we measure a cavity-field decay rate of $\kappa_D = (120 \pm 30) \:$GHz resulting in a quality factor of $Q_{exp-D} = (3900\: \pm \: 700) $.
This value agrees with $Q_{exp-C}$ (table \ref{tab:summary}) reported in the main manuscript, confirming both the results and the degradation of the cavity due to vibrations. Taking into account the cavity $Q_{exp-D}$ and the position of the nanodiamond within the hemispherical cavity structure, we estimate a maximum achievable Purcell factor ($F_{vib}$), under experimental conditions, of 12. From the ratio between $F_p$ and $F_{vib}$ we estimate an alignment of the D transition with the electric field of the cavity mode of 31\% (equation \ref{eq:detuning}). This includes both the dipole misalignment and the reduction in effective coupling strength due to the branching ratio.

\vspace{10 pt}
\section{Coupling to higher-order mode}\label{higher_mode}

\begin{figure}[ht]
\includegraphics[]{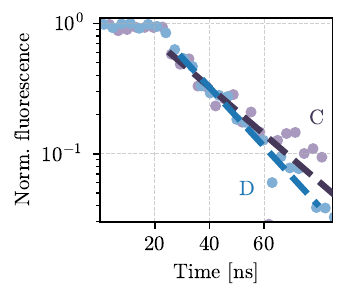}
    \caption{Pulsed lifetime measure of transition C and D coupled to an higher-order cavity mode. The resulting lifetime values from the linear fit are ($23 \: \pm \: 3$) ns and ($22 \: \pm \: 2$) ns, respectively.}
    \label{fig:highermode}
\end{figure}

As shown in Figure \ref{fig:coupling}b), the \SnV center, for a cavity length of \SI{3.4}{\micro\meter}, couples to a higher-order cavity mode. 

To investigate whether this coupling leads to a Purcell enhancement of the spontaneous emission rate for optical transitions C (or D), the OPO laser wavelength is fixed to excite transition C (D) through the cavity mode $m_{exc} = 13$ and detect ZPL counts through such higher-order mode. Achieving this double-resonance condition requires setting the effective cavity length to $L_{eff}$=\SI{3.4}{\micro\meter}. 
As in \ref{par:lifetime}, the OPO is pulsed off with an EOM while recording the fluorescence decay rate. During the acquisition time, the cavity is kept in resonance with the emitter frequency by actively adjusting the cavity length to compensate for temperature-induced drifts during the measuring time. The resulting histogram is fitted with a single exponential function yielding lifetime values of ($23 \: \pm \: 3$) ns for optical transition C and ($22 \: \pm \: 2$) ns for D (figure \ref{fig:highermode}). The larger uncertainties arise from the lower number of detected photons due to reduced coupling efficiency to the higher-order mode. Comparing these lifetimes with the free-space lifetimes reported in table \ref{free_space} shows no measurable reduction, indicating the absence of significant Purcell enhancement in this configuration. Although the emitter is coupled to a higher-order cavity mode at shorter cavity length, the coupling strength remains too weak to substantially modify the spontaneous emission rate. As a result, the cavity does not noticeably affect the emitter lifetime under these conditions.

\begin{table*}[ht]
 \begin{tabular}{l l l c l l}
\hline
\hline
\textbf{Parameter}  & \textbf{Condition} & \textbf{Symbol} & \textbf{Value} & \textbf{Units} & \textbf{Origin}\\
         \hline
         \hline
         \textbf{Emitter Properties}\\
         ZPL  & 297 K & $ZPL$ & 620.1 $\pm0.1$ & nm & Gaussian fit \\
          free-space lifetime & 297 K & $\tau_0$ & 21.7 $\pm$ 0.3 & ns & Pulsed lifetime measure \\
         C Transition & \SI{100}{K} & $ZPL_{c,100K}$ & 618.6 $\pm$ 0.1 & nm & Spectrum \\
         C Transition linewidth & \SI{100}{K} & $\gamma$ & 210 $\pm$ 20 & GHz & Lorentzian fit \\
         C Transition & \SI{4}{K} & $ZPL_{c,4K}$ & 618.54 $\pm$ 0.03 & nm & Lorentzian fit \\
         D transition & \SI{100}{K} & $ZPL_{d,100K}$ & 620.3 $\pm$ 0.1 & nm & Spectrum \\
         D Transition linewidth & \SI{100}{K} & $\gamma_D$  & 410 $\pm$ 20 & GHz & Lorentzian fit \\
         D Transition & \SI{4}{K} & $ZPL_{d,4K}$ &  620.22$\pm$ 0.03 & nm & Spectrum \\
         \hline
         \textbf{Mirror}\\
         Radius of curvature x & & $ROC_x$ & 25 & \SI{}{\micro\meter} & AFM \\
         Radius of curvature y & & $ROC_y$ & 22 & \SI{}{\micro\meter} & AFM \\
         Transmission & 618 nm & $T_{618}$ & 650 & ppm & Data from Laseroptik \\
         Transmission & 533 nm & $T_{533}$ & 6400 & ppm & Data from Laseroptik \\
         \hline
         \textbf{Cavity}\\
         Theoretical Finesse & 618 nm & $F_{ideal, 618}$ & 4800 & & Data from Laseroptik \\
         Measured Finesse & 618 nm & $F_{618}$ & 4600$\pm$ 500 & & Cavity scan \\
        Theoretical Finesse & 533 nm& $F_{ideal, 533}$ & 400 & & Data from Laseroptik \\
         Measured Finesse & 533 nm  & $F_{533}$ & 400 $\pm$ 100  & & Cavity scan \\
          Theoretical Quality factor &  & $Q_{ideal}$ & 56400 &  & $F_{618}$, $m_{det}$  \\
           Beam waste &  & $\omega_{0}$ & 1.31 & \SI{}{\micro\meter} & ROC, $L_{eff}$ \\
         Mode volume & & $V_{mode}$ & 21 & $\lambda^3$ & $\omega_{0}$, $L_{eff}$ \\
         Ideal Purcell & & $F_{cav}$ & 205 &  & $Q_{ideal}$, $\omega_{0}$, $V_{mode}$ \\
         Measured Linewidth & & $\kappa$ & 15 & GHz &  Lorentzian fit \\
           \hline
         \textbf{Measuring cavity parameters}\\
         Effective cavity length & & $L_{eff}$& 3.7 & \SI{}{\micro\meter} & WLED spectrum \\
         Detection mode number  & & $m_{det}$ & 12 & & WLED spectrum, m \\
         Excitation mode number  & & $m_{exc}$ & 14 & & WLED spectrum \\
        \hline
         \textbf{Purcell enhancement C transition}\\
         Measured Quality factor & \SI{4}{K} & $Q_{exp}$ & 3100 $\pm$ 600 &  &  Lorentzian Fit\\
         Vibration-limited Purcell & \SI{4}{K} & $F_{vib}$ & $10 \pm 2 $ &  &  $Q_{exp}$\\
         Measured Purcell & \SI{4}{K} & $F_{p}$ & 1.78 $\pm$ 0.04 &  & equation \ref{eq:Purcell}\\
         Measured Purcell & 40 K & $F_{p, 40K}$ & 1.55 $\pm$ 0.04 &  & equation \ref{eq:Purcell}\\
         Cavity-Emitter lifetime  & \SI{4}{K} & $\tau_{4K}$ & 12.2 $\pm$ 0.3 & ns & Pulsed lifetime measure \\ 
         Cavity-Emitter lifetime  & 40 K & $\tau_{40K}$ & 15.8 $\pm$ 0.2 & ns & Pulsed lifetime measure \\ 
         Cavity-Emitter lifetime  & \SI{100}{K} & $\tau_{100K}$ & 21 $\pm$ 1 & ns & Pulsed lifetime measure \\
          \hline
         \textbf{Purcell enhancement D transition}\\
         Measured Quality factor & \SI{4}{K} & $Q_{exp-D}$ & 3900 $\pm$ 700 &  &  Lorentzian Fit\\
         Vibration-limited Purcell & \SI{4}{K} & $F_{vib}$ & $12 $ &  &  $Q_{exp}$\\
         Measured Purcell & \SI{4}{K} & $F_{p}$ & 1.67 &  & equation \ref{eq:Purcell}\\
         Cavity-Emitter lifetime  & \SI{4}{K} & $\tau_{4K}$ & 13.1 $\pm$ 0.3 & ns & Pulsed lifetime measure \\ 
         Cavity-Emitter lifetime  & \SI{100}{K} & $\tau_{100K}$ & 21 $\pm$ 2 & ns & Pulsed lifetime measure \\
     \hline
       \end{tabular}
    \caption{Summary table of the main parameters.}
    \label{tab:summary}
\end{table*}

\bibliography{bibliography}

\end{document}